\documentclass[12pt]{article}
\usepackage{graphicx,floatflt}
\date{}
\newcommand{\be}{\begin{equation}}
\newcommand{\ee}{\end{equation}}
\newcommand{\bea}{\begin{eqnarray}}
\newcommand{\eea}{\end{eqnarray}}
\newcommand{\bear}{\begin{equation}\begin{array}}

\newcommand{\bm}{\boldmath}
\newcommand{\fr}[2]{\frac{{\displaystyle #1}}{{\displaystyle #2}}}

\def\lsi{\raise0.3ex\hbox{$<$\kern-0.75em\raise-1.1ex\hbox{$\sim$}}}
\def\gsi{\raise0.3ex\hbox{$>$\kern-0.75em\raise-1.1ex\hbox{$\sim$}}}
\newcommand{\lsim}{\mathop{\lsi}}

\newcommand{\epe}{\mbox{$e^+e^-\,$}}
\newcommand{\ggam}{\mbox{$\gamma\gamma\,$}}
\newcommand{\egam}{\mbox{$e\gamma\,$}}
\newenvironment{Itemize}{\begin{list}{$\bullet$}%
{\setlength{\topsep}{0.2mm}\setlength{\partopsep}{0.2mm}%
\setlength{\itemsep}{0.2mm}\setlength{\parsep}{0.2mm}}}%
{\end{list}}
\newcounter{enumct}

\title {Effective Photon Spectra for the Photon Colliders}

\author{I.F. Ginzburg\\
 Institute of Mathematics,
Prosp.~Acad.~Koptyug 4, 630090 Novosibirsk, Russia,\\
 e-mail: ginzburg@math.nsc.ru\,,\\
 G.L. Kotkin\\
Novosibirsk State University, Ul.~Pirogova 2, 630090 Novosibirsk,
Russia,\\ e-mail: kotkin@math.nsc.ru}

\begin{document}

\maketitle
\begin{abstract}
The luminosity distribution in the effective $\gamma\gamma$ mass
at a photon collider has usually two peaks which are well
separated: the High energy peak with mean energy spread about
$5-7\%$ and the wide low energy peak. The low energy peak depends
strongly on details of design and is unsuitable for the study of
New Physics phenomena. We find simple approximate form of spectra
of {\em colliding photons} for \ggam\ and \egam\ colliders,whose
convolution describes high energy luminosity peak with good
accuracy in the most essential preferable region of parameters.
\end{abstract}

\section{Introduction}

The photon colliders (\ggam and \egam) were proposed and discussed
in details in Refs.~\cite{GKST}. The forthcoming papers
\cite{modern,telnov} present some new details of design and
analysis of some effects involved in conversion.

In a basic scheme two electron beams leave the final focus system
and travel towards the interaction point (IP). At the conversion
point (CP) at the distance $b\sim 1-10$~mm before IP they collide
with focused laser beams. The Compton scattering of a laser photon
on an electron produces a high energy photon. The longitudinal
motion of this photon originates from that of an electron, so that
these photons follow the trajectories of electrons (to the IP)
with additional angular spread $\sim 1/\gamma$. With reasonable
laser parameters, one can "convert" most of the electrons into the
high energy photons. Without taking into account rescattering of
electrons on the next laser photons, the total \ggam\ and \egam\
luminosities are ${\cal L}_{\gamma\gamma}^0= k^2{\cal L}_{ee}$ and
${\cal L}_{e\gamma}^0= k{\cal L}_{ee}$ where $k$ is the conversion
coefficient and ${\cal L}_{ee}$ is the geometrical luminosity of
basic $ee$ collisions, which can be made much larger than the
luminosity of the basic \epe\ collider. Below we assume distances
$b$ and the form of the electron beams to be identical for both
beams.

Let the energy of the initial electron, laser photon and high
energy photon be $E$, $\omega_0$ and $\omega$. We define as usual
\be
x=\fr{4E\omega_0}{m_e^2}\,,\quad y=\fr{\omega}{E}\leq
y_m=\fr{x}{x+1}\,. \label{xdef} \ee The quality of the photon
spectra is better for higher $x$. However at
$x>2(1+\sqrt{2})\approx 4.8$ the high energy photons can disappear
via production of \epe pair in its collision with a following
laser photon. That is why the preferable conversion is at $x=4-5$.

The energies of colliding photons $\omega_i=y_iE$ can be
determined for each event by measuring the total energy of the
produced system $\omega_1+\omega_2$ and its total (longitudinal)
momentum $\omega_1 - \omega_2$. We discuss in more detail the main
area for study of New Physics --- {\bf the high energy region}
where energies of both photons are large enough. For the
definiteness we consider the photon energy region
$(y_m/2)<y_i<y_m$ and demand additionally that no photons with
lower energy contribute to the entire distribution over the
effective mass of \ggam\ system $2zE$ or its total energy $YE$:
\be
\fr{y_m}{2}<y_1,\,y_2<y_m\,\Rightarrow
\left(\fr{y_m}{\sqrt{2}}<z=\sqrt{y_1y_2}<y_m\right)\,,\quad
\left(1.5y_m<Y=y_1+y_2<2y_m\right)\,. \label{area}
\ee
In the interesting cases this choice covers the high energy peak in
luminosity since the photon spectra are concentrated in the more
narrow regions near $y_m$.

The growth of the distance $b$ between IP and CP is accompanied by
two phenomena. First, high energy collisions become more
monochromatic. The high energy part of luminosity is concentrated
in a relatively narrow peak which is separated well from an
additional low energy peak. This separation becomes stronger at
higher $x$ and $b$ values.  Second, the luminosity in the high
energy region decreases (relatively slowly at small $b$ and as
$b^{-2}$ at large $b$). Only the high energy peak is the area for
study of the New Physics phenomena. The low energy peak is the
source of background at these studies. The separation between
peaks is very useful to eliminate background from the data.
Therefore, some intermediate value of $b$ provides the best
conditions for study of New Physics.

Let us discuss spectra neglecting rescattering, for beginning. At
$b=0$ the \ggam luminosity distribution is a simple convolution of
two photon spectra of separate photons. At $b\neq 0$ the
luminosity distribution is a more complicated convolution of the
above photon spectra with some factor depending on $b$ and the
form of initial electron beams. With the growth of conversion
coefficient the effect of rescattering of electrons on laser
photons enhances and make this distribution dependent on the
details of design (mainly, in low energy part).

In this paper we continue the discussion from Ref.~\cite{GKST}
about main parameters of scheme which are preferable for the \ggam
and \egam colliders for the elliptic electron beams. We find the
universal description of high energy peak in this preferable
region of parameters. It allows us to obtain the remarkable
approximate form of spectra of {\bf colliding photons}, whose
simple convolution describes high energy luminosity peak with
reasonable accuracy.

\section{ Luminosity distribution without rescattering. Elliptic
electron beams}

The high energy peak in luminosity is described mainly by a single
collision of an electron with a laser photon, this part of
distribution depends on the form of initial beams only. Therefore
we start with the detailed discussion of effects from single
electron and laser photon collision. At first, we repeat some
basic points from Refs.~\cite{GKST}.

The scattering angle of produced photon $\theta$ is related to its
energy as
\be
\theta=\fr{g(x,y)}{\gamma}\,,\quad g(x,y)=\sqrt{\fr{x}{y}-x-1}\,,
\quad \gamma=\fr{m_e}{E}\,.\label{angle}
\ee

Let the mean helicities of initial electron, laser photon and high
energy photon be $\lambda_e/2$, $P_\ell$ and $\xi_2$. The energy
spectrum of produced photons is ($N$ is normalization factor)
\be
F(x,y)=N\left[\fr{1}{1-y}-y+(2r-1)^2 -\lambda_eP_\ell\;
xr(2r-1)(2-y)\right]\,, \quad r=\fr{y}{x(1-y)}\,. \label{spectr}
\ee At $\lambda_eP_\ell =-1$ and $x>1$ this spectrum has a sharp
peak at high energy which becomes even more sharp with $x$ growth.

The spectrum is more sharp when $-\lambda_eP_\ell$ is larger. We
present below mainly the values from the real projects $\lambda_e=
0.85$, $P_\ell=-1$.

The degree of circular polarization of high energy photon is
\be
<\xi_2>=N\fr{\lambda_e xr\left[1+(1-y)(2r-1)^2\right]
-P_\ell\;(2r-1)\left[\fr{1}{1-y}+1-y\right]}{F(x,y)}\,.
\label{polariz}
\ee

The photons with lower energy have higher production angle
(\ref{angle}). With the growth of $b$, these photons spread more
and more, and they collide only rarely. Therefore, with the growth
of $b$ photon collisions become more monochromatic, only high
energy photons taking part in these collisions the low energy part
of total luminosity being rejected (here photons in the average
are almost nonpolarized).

This effect was studied in \cite{GKST} for the gaussian round
electron bunches. However, the incident electron beams are
expected to be of an elliptic form with large enough ellipticity.

Let initial electron beams be of the gaussian elliptic form with
vertical and horizontal sizes $\sigma_{ye}$ and $\sigma_{xe}$
atthe IP (calculated for the case without conversion). The
discussed phenomena are described by a reduced distance between
conversion and collision points $\rho$ and an aspect ratio $A$
\be
\rho^2=\left(\fr{b}{\gamma\sigma_{xe}}\right)^2+
\left(\fr{b}{\gamma\sigma_{ye}}\right)^2\,,\quad
A=\fr{\sigma_{xe}}{\sigma_{ye}}\,.\label{rho}
\ee
The luminosity distributions in this case can be calculated by the
same approach which was used in Ref.~\cite{GKST}.\\

The distribution of the photons colliding with opposite electrons
at {\bf\bm \egam collider} is
\be
\fr{d{\cal L}_{e\gamma}}{dy} =\int\fr{d\phi}{2\pi}F(x,y)\exp\left[
- \fr{\rho^2g^2(x,y)}{4(1+A^2)}(A^2\cos^2\phi +\sin^2\phi)\right]\,.
\label{egamspectr}
 \ee
For the round beams ($A=1$) we have in the exponent
$\rho^2g^2(x,y)/8$.\\

The distribution of the colliding photons over their energies at
{\bf\bm \ggam collider} is \bear{c}
 \fr{d^2{\cal L}_{\gamma\gamma}}{dy_1dy_2}
=\int\fr{d\phi_1d\phi_2}{(2\pi)^2}F(x,y_1)F(x,y_2)\exp\left[
-\fr{\rho^2\Psi}{4(1+A^2)}\right]\,,\\
\\
\Psi=
A^2\left[g(x,y_1)\cos\phi_1+g(x,y_2)\cos\phi_2\right]^2 +
\left[g(x,y_1)\sin\phi_1+g(x,y_2)\sin\phi_2\right]^2\,.
\end{array}\label{ggamspectr}
\ee For the round beams ($A=1$) one can perform integrations over
$\phi_i$ in the analytical form. It results in the equation from
Ref.~\cite{GKST} with the Bessel function of an imaginary argument
$I_0(v^2)$ \bear{c} \fr{d^2{\cal
L}_{\gamma\gamma}}{dy_1dy_2}=F(x,y_1)F(x,y_2)\exp\left[-
\fr{\rho^2}{8}(g^2(x,y_1)+g^2(x,y_2))\right]I_0(v)\,,\\
v^2=\fr{\rho^2}{4}g(x,y_1)g(x,y_2)\,.
\end{array}\label{round}
\ee

We have analyzed numerically the high energy part of the
luminosity (in the region (\ref{area})) ${\cal L}^h$ as a function
of $\rho^2$ and $A$ at $2<x<5$, $-\lambda_eP_\ell\geq 0$. (We use
this notation for both total luminosity integrated over the region
(\ref{area}) and differential distributions.)

The growth of $\rho$ results both in a better form of luminosity
distribution and reduction of luminosity. We find that the
luminosity ${\cal L}^h$ depends only weakly on the aspect ratio
$A$ at $A>1.5$ and $\rho^2<1.3$. At $\rho^2\leq 1$ this dependence
is weak at all values of $A$ (including $A=1$). For
$\lambda_eP_\ell=-0.85$ the luminosity ${\cal L}^h$ at $\rho=1$
contains large enough fraction from the high energy part of
luminosity given at $\rho=0$.  For the unpolarized case
($\lambda_eP_\ell=0$) both this fraction is smaller and the high
energy peak is separated weakly from the low energy one.  Some of
these statements can be seen from the table below where we show
the ratio of high energy luminosity ${\cal L}^h$ to the total
luminosity ${\cal L}_{\gamma\gamma}^0$ at some values of
parameters.
\begin{center}
\begin{tabular}{|c|c|c||c|c|}\hline
&\multicolumn{2}{|c||}{$\rho=0$, any $A$
}&\multicolumn{2}{|c|}{$\rho=1$, $A\geq 1.5$}\\\hline
$\lambda_eP_\ell$&-0.85&0&-0.85&0\\\hline\hline
$x=4.8$&0.35&0.25&0.28&0.19\\\hline
$x=2$&0.29&0.19&0.25&0.16\\\hline
\end{tabular}
\end{center}
It seems unreasonable to use lowst part of total luminosity than
that obtained at $\rho=1$.

To have a more detailed picture for the simulation, we study the
luminosity distribution in the relative values of the effective
mass $z=W/(2E)$ and the total energy $Y={\cal E}/E$ of the pair of
colliding photons (\ref{area}). Their typical forms for different
values of aspect ratio $A$ are shown in Figs.~1,~2.

\begin{figure}[hb]
\includegraphics[width=0.47\textwidth,height=7cm]{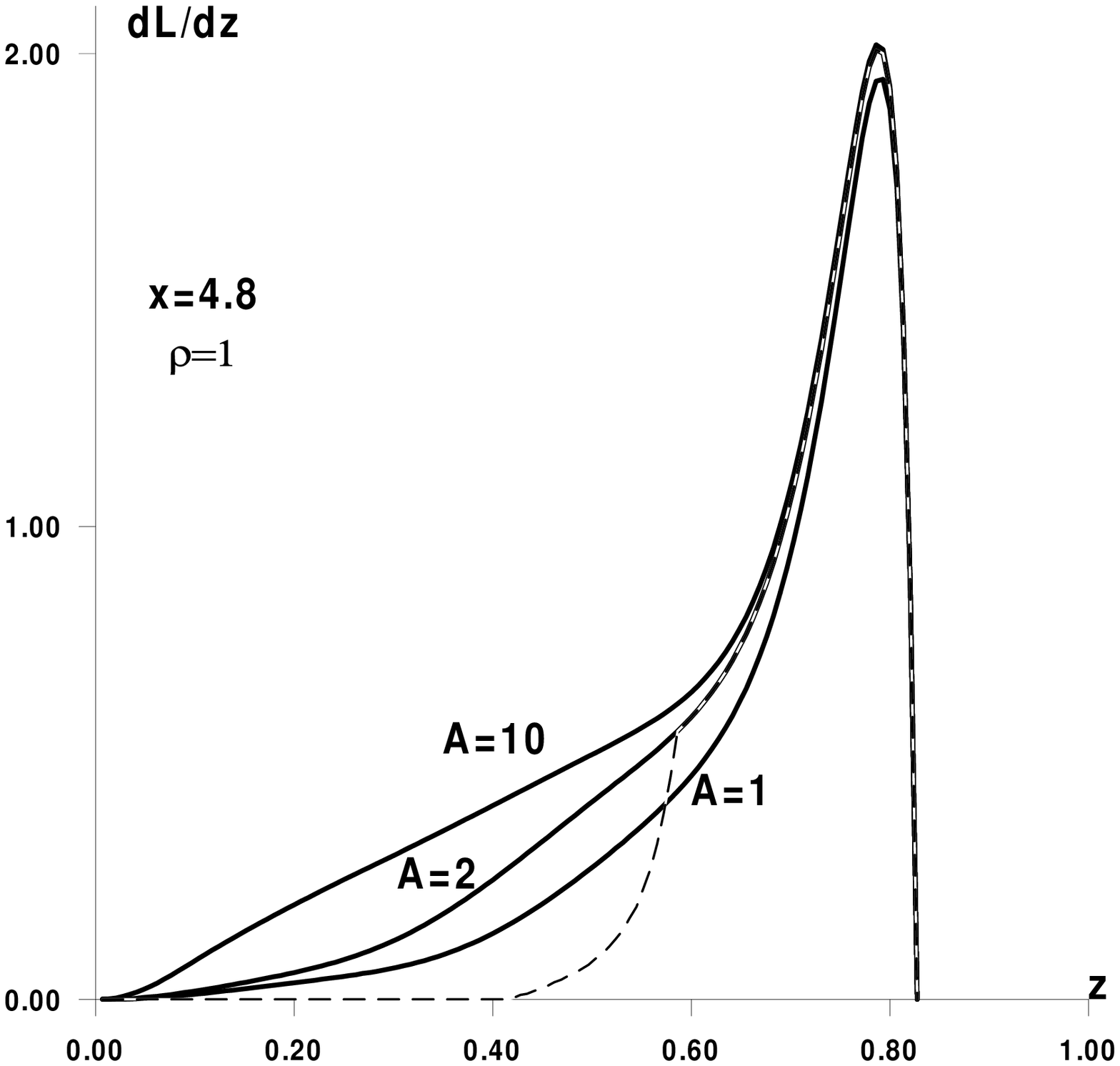}
\hfill
   \includegraphics[width=0.47\textwidth,height=7cm]{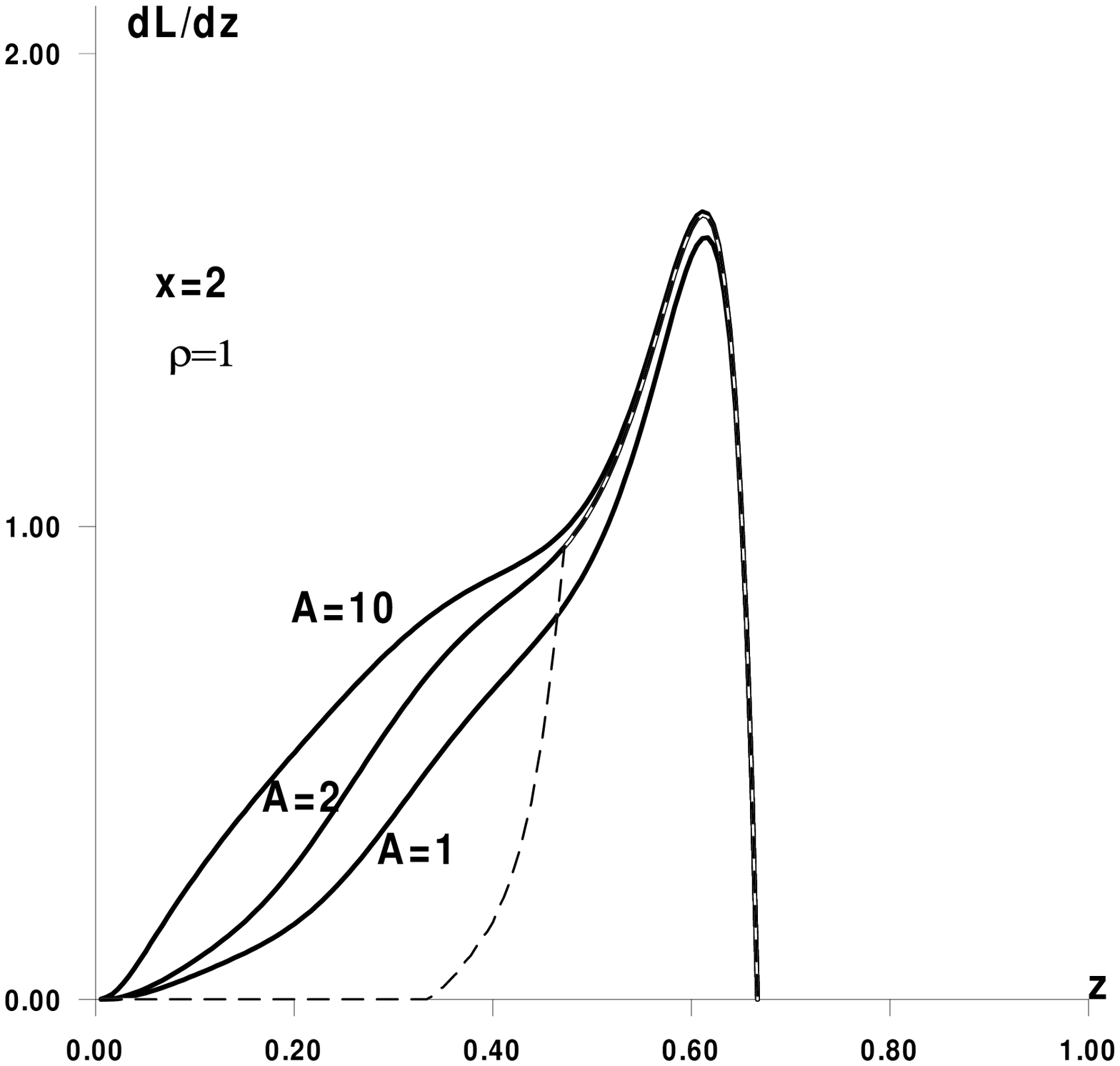}
\caption{\em The luminosity distributions in the effective mass of
\ggam\ system at $\lambda_eP_\ell=-0.85$. Dashed line presents this
 distribution for $y_i>y_m/2$ at $A=2$.}
\label{figappr3}
\end{figure}

\begin{figure}[htb]
\includegraphics[width=0.47\textwidth,height=7cm]{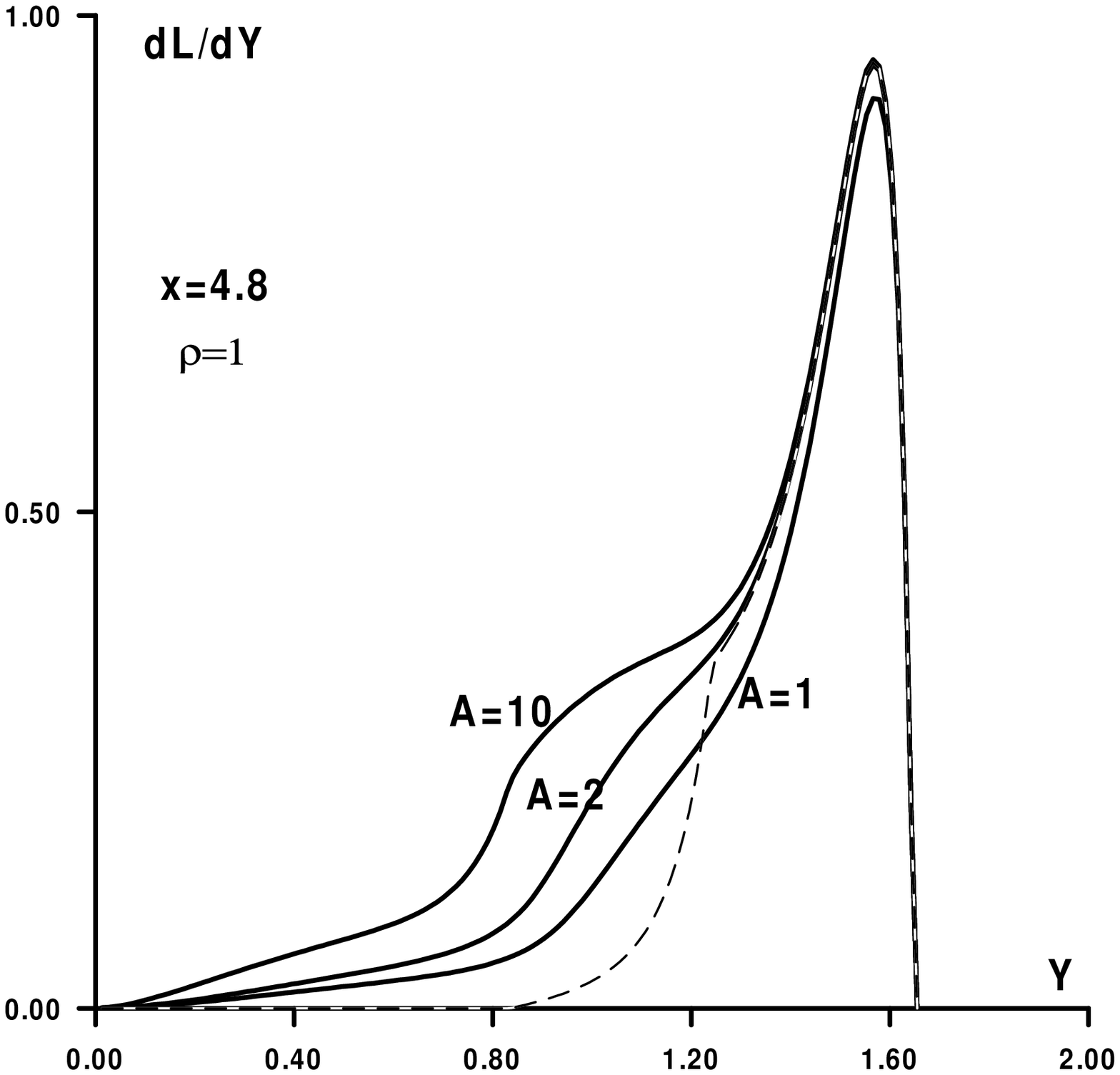}
 \hfill
   \includegraphics[width=0.47\textwidth,height=7cm]{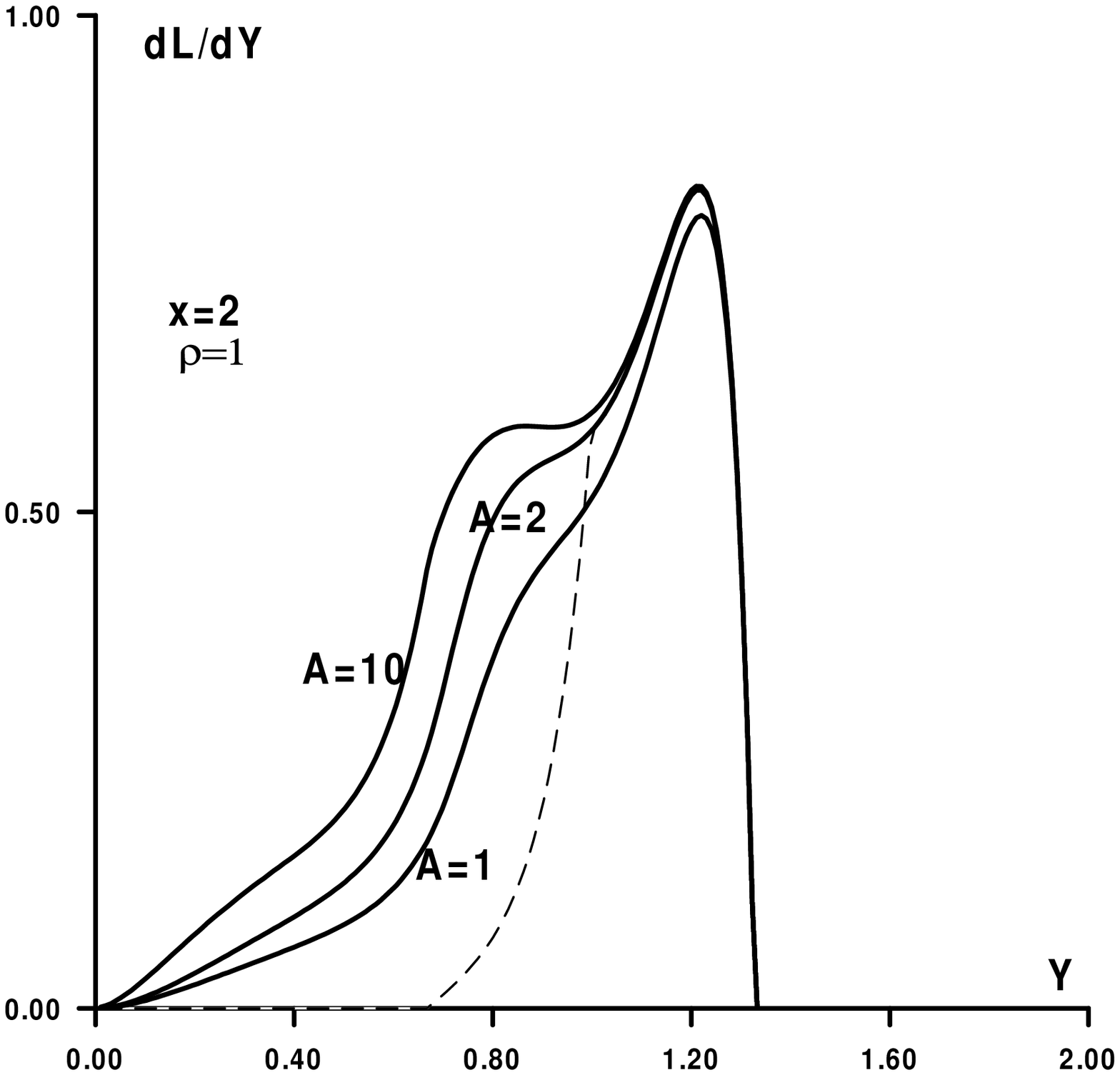}
\caption{\em The luminosity distributions in the total energy of
\ggam\ system at $\lambda_eP_\ell=-0.85$. Dashed line
presents this distribution for $y_i>y_m/2$ at $A=2$.}
 \label{figappr4}
\end{figure}

The numerical study of these distributions shows us that its high
energy part is practically the same for all values $A>1.5$ at
fixed $\rho^2< 1.3$ (with small difference near lower part of
peak). The luminosity within high energy peak for round beams
($A=1$) is slightly lower than that for the elliptic beams. This
difference is about 5\% in the main part of region below peak. At
$\rho^2=1$ this difference is small for all $A$. At $\rho^2>0.5$
the high energy part of luminosity has the form of a narrow enough
peak. This peak is not so sharp at lower $x$ and it is even less
sharp at $\lambda_e P_\ell=0$.

With the growth of aspect ratio $A$ the entire distributions
acquire low energy tails (as compared with round beams) originated
from the collisions of low energy photons scattered near the
horizontal direction with the opposite high energy photon
scattered in the vertical direction. This tail is added to that
from the rescatterings and is not of much  interest in the
discussion of high energy peak. At higher $\rho$ and $A$ this
effect becomes more essential in the region of the peak.

As a result, the preferable region of parameters for photon colliders
is
\be
5>x>2\,,\quad -\lambda_eP_\ell\,\geq 0.5\,,\quad \rho^2<1.3\,.
\label{region}
 \ee
 Additionally, here the high energy peaks in the \ggam and \egam
 luminosities  are described by one parameter $\rho$ and practically
 independent from $A>1.5$.

\section{Rescattering contribution to the spectra. Qualitative
description}

The rescattering of electrons on the following laser photons
produces new high energy photons ({\em secondary photons}) which
modify the luminosity distribution mainly in the low energy part.
Detailed form of additional components of luminosity distribution
depends strongly on the conversion coefficient and other details
of design. That is why we present here only qualitative discussion
with some simple examples.

Let us enumerate the differences in properties of secondary photons
from those from the first collision (we will denote them as {\em
primary photons}).
\newline
(1) The energy of secondary photons is lower than that of primary
photons.
\newline
(2) There is no definite relation between energy and production angle
like (\ref{angle}).
\newline
(3) The mean polarization of secondary photons is practically zero.

Fig.~3 presents a typical energy spectrum of photons with only one
rescattering at conversion coefficient $k=1$. Dashed line
represents fraction of secondary photons. Let us explain some
features of this spectrum.

\begin{figure}[htb]
   \centering
\includegraphics[width=10cm,height=7cm]{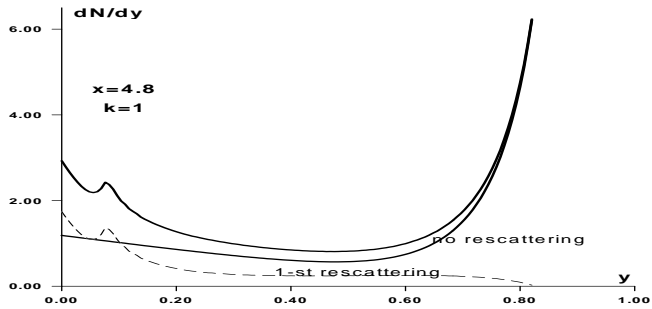}
\caption{\em The energy spectrum of photons with one rescattering,
$\lambda_eP_\ell=-0.85$.}
   \label{figggam}
\end{figure}

The main fraction of electrons after the first scattering has the
energy which corresponds to the peak in produced photons spectra,
$E_e \approx E-\omega_m=E/(x+1)$. For the next collision $x\to
x/(x+1)$. Therefore, the additional energy peak in photon spectrum
caused by secondary photons from the first rescattering is at the
photon energy $\sim y_m/(2x+1)$, it is much lower than $y_mE$ for
$x>2$. The subsequent rescatterings add more soft photons.
Besides, the primary photon spectrum (\ref{spectr}) is
concentrated near its high energy boundary. So, the fraction of
scattered electrons with energy close to $E$ is small, the effect
of rescattering in a high energy part on the entire photon
spectrum is also small. In the subsequent rescatterings such peaks
become smooth. The well known result is a large peak near $y=0$.
Note also that the secondary photons in average are nonpolarized.

The shape of an additional contribution of secondary photons to
the luminosity distributions (second--second and primary--second)
depends on $k$, $\rho$ and $A$. Nevertheless, the different
simulations show the common features (see \cite{modern,telnov}):

At $\rho^2\geq 0.5$, $k\lsim 1$ the luminosity distribution has
two well separated peaks: {\bf high energy peak (mainly from
primary photons) and wide low energy peak (mainly from secondary
photons)}. Photons in the high energy peak have high degree of
polarization, mean polarization of photons in the low energy peak
is close to 0. At smaller $x$ or $-\lambda_eP_\ell$ this
separation of peaks becomes less definite and mean photon
polarization becomes less then that given by Eq.~(\ref{polariz}).

With a good separation of peaks, the backgrounds from the low
energy peak could be eliminated relatively easily in many
problems.

\section{Approximation}

The previous discussion shows us that there are chances to
construct an approximation for photon spectra which would describe
the high energy peak simply and in the universal way. We were
searching for an approximation in which the high energy peak in
\ggam luminosity would be given by a simple convolution of form
\be
\fr{d^2{\cal L}}{dy_1dy_2}= F_a(x,y_1,\rho^2)F_a(x,y_2,\rho^2)\,.
\label{appr1} \ee instead of complex integration
(\ref{ggamspectr}) (independent from aspect ratio $A$).

We tested different forms of effective photon spectra. Taking into
account form of angular spread of separate beam, we consider a
test function for the high energy peak in the form
\be
F_a(x,y,\rho^2) =\left\{\begin{array}{cc}
 F(x,y)\exp\left[ -B\rho^2 g(x,y)^2/8\right]&\mbox{ at }y>y_m/2\,,\\
 0 &\mbox{ at } y<y_m/2\,.\end{array}\right.
\label{appr2}
\ee
where coefficient $B$ is varied.

A good fit for the high energy peak at $2<x<5$, $\rho^2<1.3$,
$A>1.5$ is given by the values \bear{ll}
 B=1&\mbox{ for the \ggam\  collider}\,,\\
B=0.7&\mbox{ for \egam\  collider}\,.
\end{array}\label{appr3}
\ee The curves in Figs.~4,5 show the accuracy of this
approximation for the distributions in both the effective \ggam
mass $z=\sqrt{y_1y_2}$ and the total photon energy $Y=y_1+y_2$ at
$\lambda_eP_\ell=-0.85$. These curves show the excellent quality
of our approximation. The distributions calculated without angular
spread factor (at $B=0$) are also shown here by dashed lines.
These curves are markedly higher than the precise curve and they
are wider than the real distributions. The first inaccuracy can be
compensated by a suitable renormalization, but the second
inaccuracy cannot be eliminated from calculations.

\begin{figure}[htb]
 \includegraphics[width=0.47\textwidth,height=7cm]{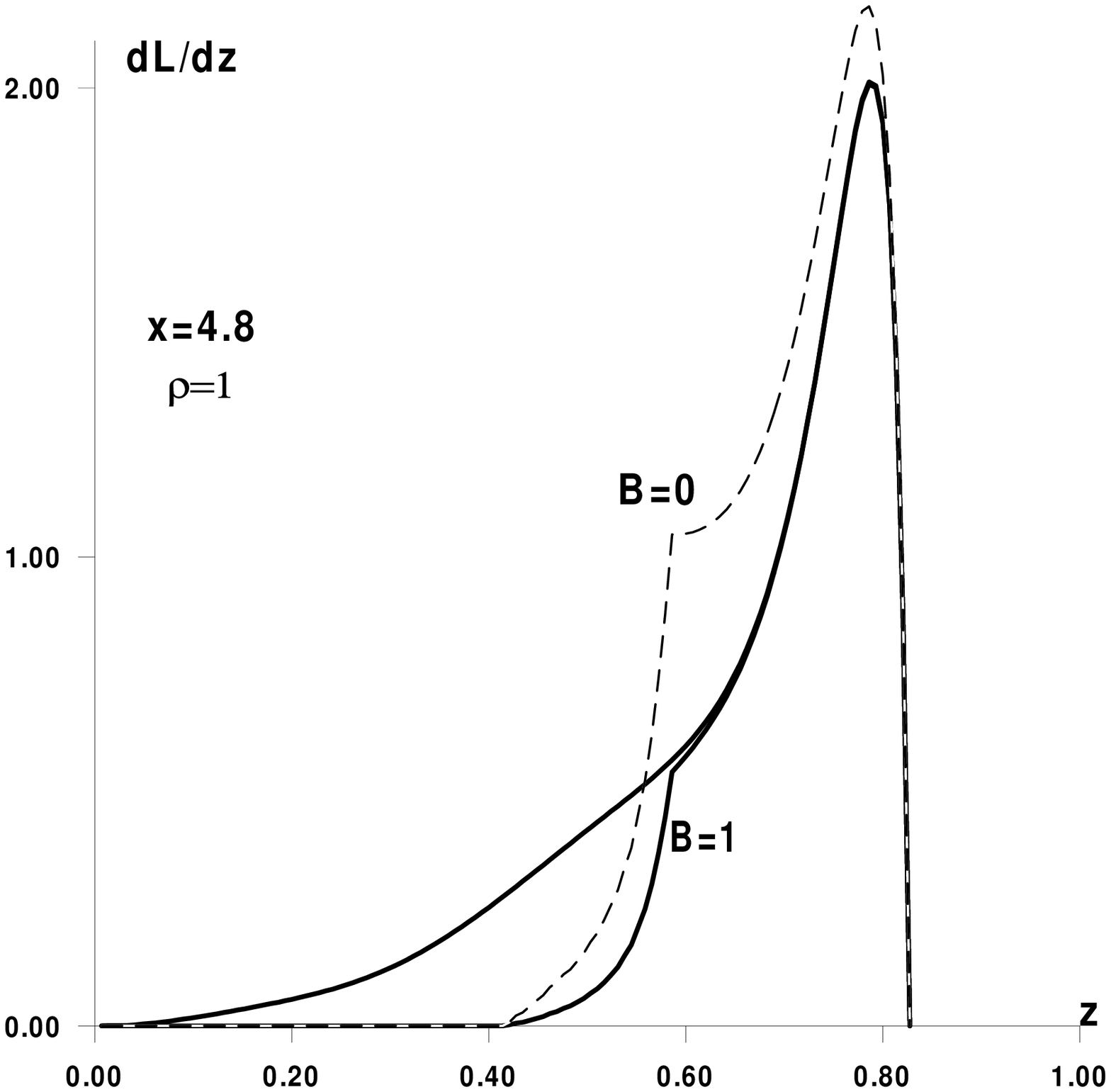}
\hfill
   \includegraphics[width=0.47\textwidth,height=7cm]{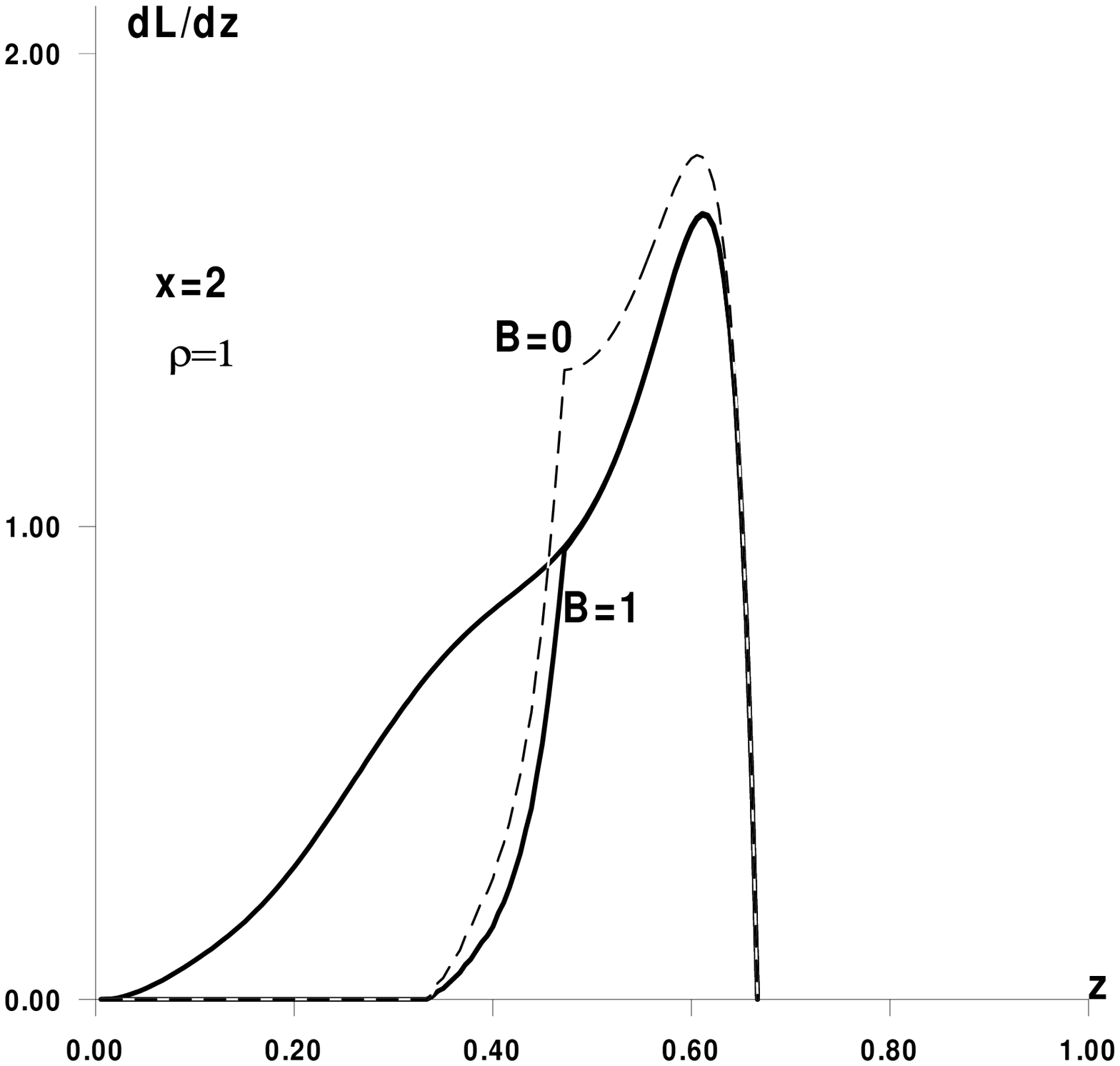}
\caption{\em The precise luminosity distribution in the effective
mass of \ggam\ system without rescattering at $A=2$ and approximation
(\ref{appr1}) -- (\ref{appr3}). Dashed line -- approximation used
spectra at $\rho=0$ in the region (\ref{area}).} \label{apprzap}
\end{figure}

\begin{figure}[htb]
 \includegraphics[width=0.47\textwidth,height=7cm]{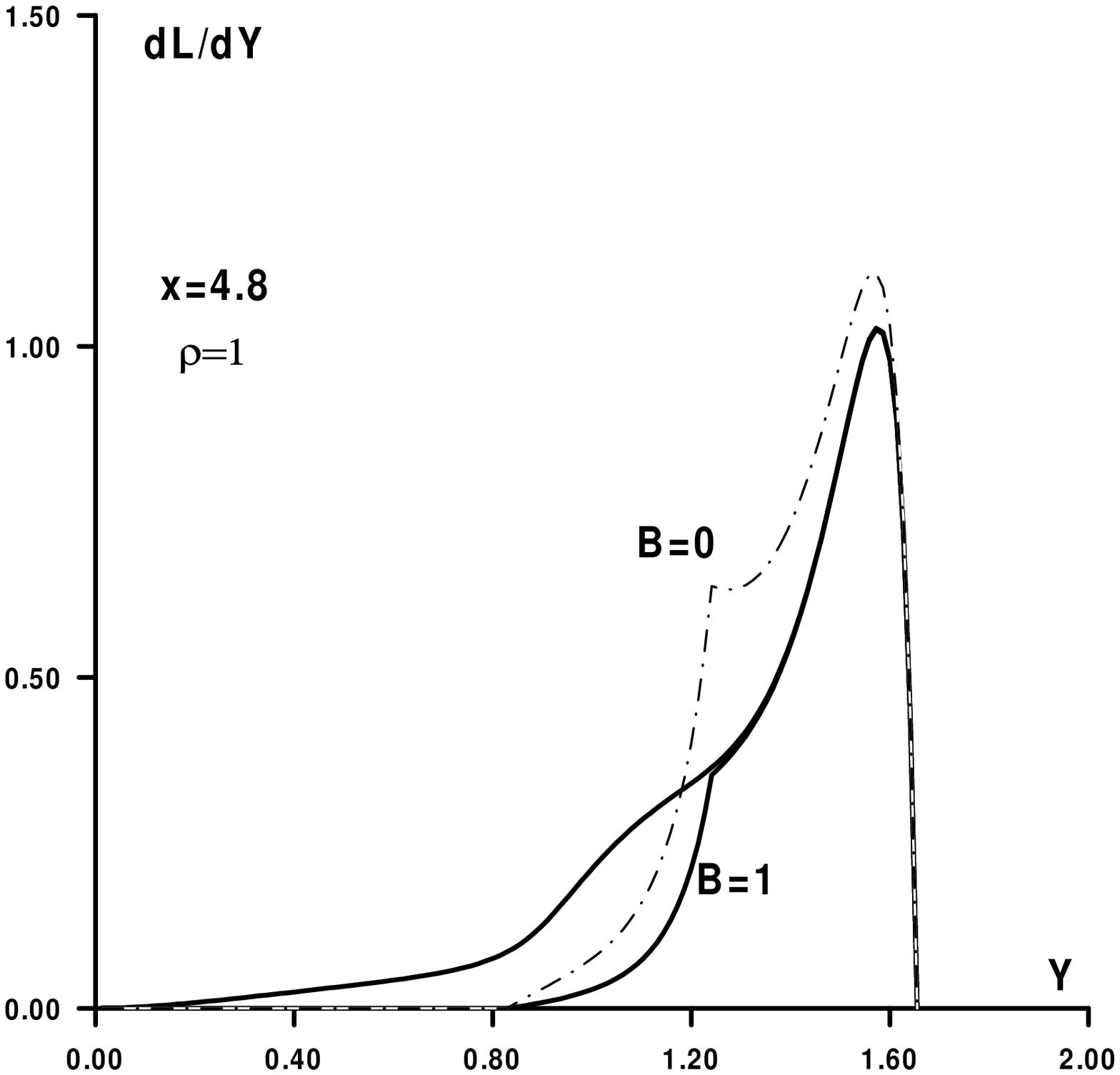}
\hfill
   \includegraphics[width=0.47\textwidth,height=7cm]{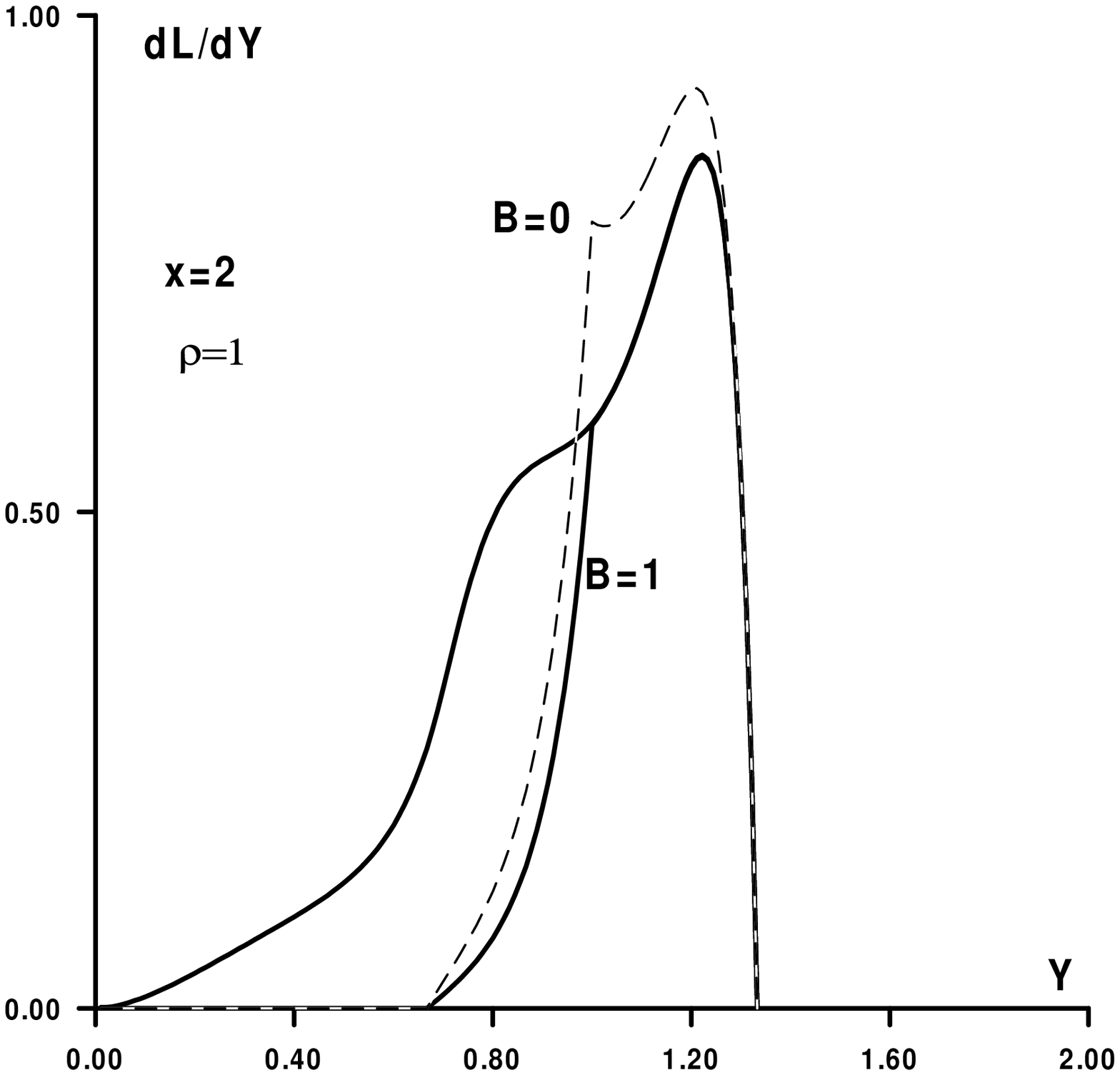}
\caption{\em The precise luminosity distribution in the total energy
of \ggam\ system without rescattering at $A=2$ and approximation
(\ref{appr1}) -- (\ref{appr3}). Dashed line -- approximation used
spectra at $\rho=0$ in the region (\ref{area}).}
\label{apprEap}
\end{figure}

Note that the approximation (\ref{appr3}) for the \ggam\ collision
can be obtained from eq.~(\ref{round}) if Bessel function
$I_0(v^2)$ are replaced by unity. Figs.~4,5 show that our
approximation coincide practically with precise distributions
within high energy peak for the elliptic beams ($A\geq 1.5$).
Therefore, the difference between curves for $A=1$ and $A=2$ in
Figs.~1,2 in the region (\ref{area}) is caused by the Bessel
function factor.

Using of "precise" Eq.~(\ref{ggamspectr}) instead of our
approximation is only a sham improvement. The difference between
the approximation and the "precise" formula is usually smaller
than the effect of rescatterings.

\section{Results}

Let us enumerate the main results.

\begin{Itemize}
\item The variable $\rho$ (\ref{rho}) is a good variable for the
description of the high energy peak in the spectral luminosity
independent from the aspect ratio $A$ at $A>1.5$, $\rho^2<1.3$,
$2<x<5$, $\lambda_eP_\ell<0$.
\item At $\rho\sim 1$ and suitable polarizations of initial beams
the high energy peak in luminosity is separated well from the low
energy peak. This separation can be destroyed by using of large
conversion coefficient or (and) values $x\approx 1$ or
$\lambda_eP_\ell>0$.
\item To discuss future experiments at photon collider with good
enough accuracy, one can use simple approximation (\ref{appr1})--
(\ref{appr3}) at $\rho=1$ instead of details simulation of
conversion and collision. In this approximation the details of
design are inessential. Possible decreasing of $\rho^2$ to the
value 0.5 can be also considered.
\item The numbers describing luminosity of Photon Collider
correspond the discussed high energy peak only.
 \end{Itemize}
\section*{Acknowledgments}

We are grateful to G. Jikia, V.G.~Serbo and V.I. Telnov for the useful discussions. This
work was supported by grant RFBR 99-02-17211 and grant of
Sankt--Petersburg Center of Fundamental Sciences.

\end{document}